\documentclass[%
 reprint,
nofootinbib,
 amsmath,amssymb,
 aps,
]{revtex4-2}

\usepackage{graphicx}
\usepackage{dcolumn}
\usepackage{bm}
\usepackage{hyperref}

\usepackage{subcaption}
\usepackage{amsmath,amssymb,bm,amsthm,mathrsfs, ascmac, braket,color,stmaryrd}
\usepackage{float}
\usepackage{footmisc}
\usepackage{xcolor}

\newcommand{\dbracket}[1]{\llbracket{#1}\rrbracket}

\begin{document}

\preprint{APS/123-QED}

\title{Enforcing exact permutation and rotational symmetries\\ in the application of quantum neural network on point cloud datasets}



\author{Zhelun Li}
 \email{zhelunli@icepp.s.u-tokyo.ac.jp}
\author{Lento Nagano}%
 \email{lento@icepp.s.u-tokyo.ac.jp}
 \author{Koji Terashi}%
 \email{terashi@icepp.s.u-tokyo.ac.jp}
 
\affiliation{%
International Center for Elementary Particle Physics (ICEPP),\\
The University of Tokyo, 7-3-1 Hongo, Bunkyo-ku, Tokyo 113-0033, Japan
}%

\date{\today}

\begin{abstract}
Recent developments in the field of quantum machine learning have promoted the idea of incorporating physical symmetries in the structure of quantum circuits. A crucial milestone in this area is the realization of $S_{n}$-permutation equivariant quantum neural networks (QNN) that are equivariant under permutations of input objects. In this work, we focus on encoding the rotational symmetry of point cloud datasets into the QNN. The key insight of the approach is that all rotationally invariant functions with vector inputs are equivalent to a function with inputs of vector inner products. We provide a novel structure of QNN that is exactly invariant to both rotations and permutations, with its efficacy demonstrated numerically in the problems of two-dimensional image classifications and identifying high-energy particle decays, produced by proton-proton collisions, with the $SO(1,3)$ Lorentz symmetry.
\end{abstract}

\maketitle


\section{Introduction}
\label{sec:intro}

Recently, there has been an increasing interest in Quantum Machine Learning (QML), which is expected to have an advantage over its classical counterpart for a specific task (see e.g.~\cite{Schuld_2014,Biamonte_2017,Mangini_2021,challenges_cerezo} for review articles) in a wide range of fields including quantum chemistry, condensed-matter physics, high-energy physics (HEP).
Among variants of QMLs, the (variational) Quantum Neural Network (QNN) e.g.~\cite{Mitarai_2018_paramShift,farhi2018classification,Benedetti:2019inj, PhysRevA.101.032308, Li2021RecentAF,Ren2022ExperimentalQA} is widely used architecture.
Specifically, in the field of HEP, a number of recent papers have pioneered the application of QNN in the analysis of high energy collider datasets~\cite{Guan_2021}, including classifying new physics events using quantum annealing~\cite{annealing}, variational quantum circuits~\cite{Terashi_2021, Wu_2021, Blance_2021}, and quantum kernel algorithms~\cite{Wu_QSVM,wozniak2023quantum, Schuhmacher_2023}.

In recent years, researchers in both classical and quantum machine learning communities have started to experiment with ideas of encoding problem-specific symmetries into QNNs. In classical Machine Learning (ML), equivariant Neural Networks (NN) have been proven to be an effective method to exploit symmetries in the dataset, which often leads to substantial improvements in model performances. An example of the equivariance in NN is the algorithm known as Convolutional Neural Networks (CNNs)~\cite{CNN_lecun, CNN}, which was designed to address the translational symmetry in images. A standard CNN would operate on gridded pixel inputs of images and perform sliding window convolutions on nearby pixels such that the network could pick up patterns independent of their translational positions. This idea of convolution-based NN was also realized in QNNs by introducing quantum convolution layers made of quasi-local unitary transformations applied in a translationally invariant fashion~\cite{QCNN_Cong, Hur_2022_QCNN}.
The equivariance of this architecture, called Quantum Convolutional Neural Networks (QCNNs), has been proven extremely useful in the study of quantum many-body problems~\cite{QCNN_Cong, PhysRevB.107.L081105, Herrmann:2021tah, Nagano:2023kge,wrobel2021application, PhysRevResearch.6.023042}, especially when the Hamiltonian under investigation possesses an exact, or approximate, translational symmetry.
In the last few years, general theories of achieving equivariance in quantum circuits have been built using ansatz choices that accommodate the algebraic properties of the underlying symmetries~\cite{nguyen2022theory, Meyer_2023, Ragone:2022axl}.  One example of these developments is the introduction of the \textit{twirling method}~\cite{QNN_permutation,nguyen2022theory, Meyer_2023}, which provides a general method to encode full permutation symmetry into the QNN.
Several recent papers have designed rotationally equivariant QNN for image classification by studying the properties of discretized approximation of the rotational group~\cite{ReflectionMNIST, EQNN,west2024provably,sebastian2024image}. 
Other examples include permutation equivariance~\cite{PRXQuantum.4.020327,Zheng:2022pwx,Das:2024fwy,Mansky:2023wen,heredge2023permutation}, $SU(N)$ equivariance~\cite{Zheng:2021sml,Le:2023bkh,Li:2023vdn,East:2023nwl,Zheng:2022dmx}, and equivaraince in graph problems~\cite{Skolik:2022qwn,Mernyei:2021krm,Verdon:2019cfh,Sauvage:2022jqd}. 
Despite the success in equivariant QNN, the realization of arbitrary symmetries, especially continuous symmetries, in QNN is still considered a challenging task~\cite{nguyen2022theory,west2024provably, EQNN} in general.  Since the convolution-based method requires the model to perform convolutions across all group elements~\cite{Cohen_groupECN}, its application in datasets with continuous symmetry inevitably runs into difficulties induced by the infinite number of elements in continuous groups. Furthermore, the traditional convolution-based treatment in classical ML requires the use of complicated convolutional layers targeting a discretized approximation, often realized by Fourier transforms, of the continuous symmetry~\cite{cohen2018spherical,weiler20183d,veeling2018rotation,shimmin2021particle}.  Such treatments are proven to be very hard to compute even in classical NNs. It is still unclear how a QNN could replicate these architectures using hardware-efficient ansatz consisting of single-qubit rotations and 2-qubit entanglements. We also observe that most realizations of equivariant QNNs utilize the amplitude encoding technique, which requires complicated circuit structures to fully encode the classical data and is therefore challenging to implement on noisy intermediate-scale quantum machines~\cite{ReflectionMNIST, EQNN,west2024provably}.

In this work, we aim to design a new framework of equivariant QNNs that is compatible with near-term quantum machines. We consider an alternative approach to enforcing exact rotational symmetries in point cloud datasets, which consist of collections of data points in arbitrary dimension spaces.  To achieve the exact equivariance, we invoke Weyl's work~\cite{weyl1946classical}, which asserted that all invariant functions of the orthogonal groups~$O(N)$ with vector inputs could be expressed as functions solely dependent on the group-invariant inner products of these vectors.  This key insight enables us to perform simple data preprocessing to obtain~$O(N)$-invariant inputs to a network that is therefore guaranteed to be exactly~$O(N)$-invariant at the input level. This approach was first developed by Villar et al.~\cite{villar2023scalars} who proved that this simple method involving only pair-wise scalars could significantly improve the performance of NN. Recently, another work in the field of HEP elaborated on this idea to produce the Permutation Equivariant and Lorentz
Invariant or Covariant Aggregator Network (PELICAN)  with the full permutation and~${SO}(1,3)$ Lorentz symmetry~\cite{PELICAN}.  The PELICAN  was tested on the top-quark identification task, which is an example of real-world problems involving~${SO}(1,3)$ Lorentz and permutation symmetries. When compared with state-of-the-art models in top-quark identifications, the PELICAN was reported to have achieved a similar performance while having significantly lower model complexity. We note that this idea of enforcing~${SO}(N)$ rotational symmetry by taking inner products can be easily extended to QNNs since it does not require any specific choice of the quantum circuit ansatz. 

This paper is organized as follows. In Section \ref{sec:methods}, we discuss the theoretical aspect of the implementations of both the rotational and permutation symmetries. For the rotational symmetry, we will briefly go through the theorem that offers the theoretical guarantee for our approach. And then we proceed to work out the architecture of quantum circuits that could further enforce permutation equivariance on the rotationally invariant inputs.  In Section \ref{sec:results}, we first test this architecture in a dataset made of rotated 2D images to demonstrate its power. Then a slightly more complex version is applied to the task of particle decay classifications. Finally, we conclude our work and discuss its effectiveness, limitations, and possible future extensions in Section \ref{sec:discussion}.

\section{Methods}
\label{sec:methods}

In this Section, we will first briefly introduce the general recipe to build a QNN with trainable parameters in Section~\ref{ssec:generalCon}. Section~\ref{ssec:QNN} gives the details of the three QNN models that are investigated in this paper: The baseline QNN with no symmetry, the QNN with rotational symmetry, and the QNN with both rotational and permutation symmetries. Finally, we introduce the loss function used in the training process in Section~\ref{ssec:nonlinear}.


\subsection{General construction}
\label{ssec:generalCon}
Suppose that our data is a set of $\{\vec{x}_{i}\}$ and associated labels $\{y_i\}$.
First, the classical vectors $\vec{x}_i$ are encoded by the encoding unitary~$U_{\text{E}}$ which acts on the initial state as $U_{\text{E}}(\{\vec{x}_i\})|\psi_0\rangle$.
Then these states are processed by a QNN unitary $U_{\text{QNN}}(\theta)$ to have a final state 
\begin{equation}
    |\psi(\{\vec{x}_i\}; \theta)\rangle=
    U_{\text{QNN}}(\theta)U_{\text{E}}(\{\vec{x}_i\})|\psi_0\rangle\,.
\end{equation}

After the feature encoding and the hardware efficient ansatz, the output of the QNN, $\hat{y}$, is evaluated by measuring the expectation value of an observable, $O$, which is taken to be the product of Pauli-$Z$ operators across all qubits in this work:

\begin{align}
    \hat{y}(\{\vec{x}_i\}; \theta)
    &=
    \langle\psi(\{\vec{x}_i\}; \theta)|O|\psi(\{\vec{x}_i\}, \theta)\rangle\, \\
    &= \langle\psi(\{\vec{x}_i\}; \theta)|Z^{\otimes N_{q}} |\psi(\{\vec{x}_i\}, \theta)\rangle\,  .
\end{align}

The output layer of the QNN uses the global observable, $Z^{\otimes N_{q}}$,  which is more prone to the barren plateau phenomenon than local observables~\cite{costFunctionBP_2021,BPkernel}.  However, the global observable is fully compatible with the twirling-based permutation equivariant ansatz that will be introduced in Section~\ref{ssec:permutation}.~\footnote{One can avoid the use of global observable e.g. by using $\sum_{n}Z_{n}$ instead. We do not investigate the (potential) difference that comes from such a replacement in this paper.\label{fn:localOB}}  The output will then be compared to the truth label of the events to perform classification. The rotation angles in the ansatz act as trainable weights, the values of which are to be optimized in the learning process.\footnote{Typically, the weights are optimized via gradient descent, which relies on the computation of the first-order derivatives of the output through the parameter-shift rule~\cite{Mitarai_2018_paramShift, Schuld_2019_paramShift}. It is known that the gradient of QNN often exhibits barren plateau~\cite{McClean_2018_BP}, in which the gradient will diminish exponentially as the number of qubits increases.} In this paper, we use a gradient-free optimizer known as COBYLA~\cite{Powell:1994xno_COBYLA} to update the weights in the learning process.

\subsection{QNN architectures}
\label{ssec:QNN}
\subsubsection{Baseline model}

\begin{figure*}
    \centering
    \includegraphics[width=0.8\textwidth]{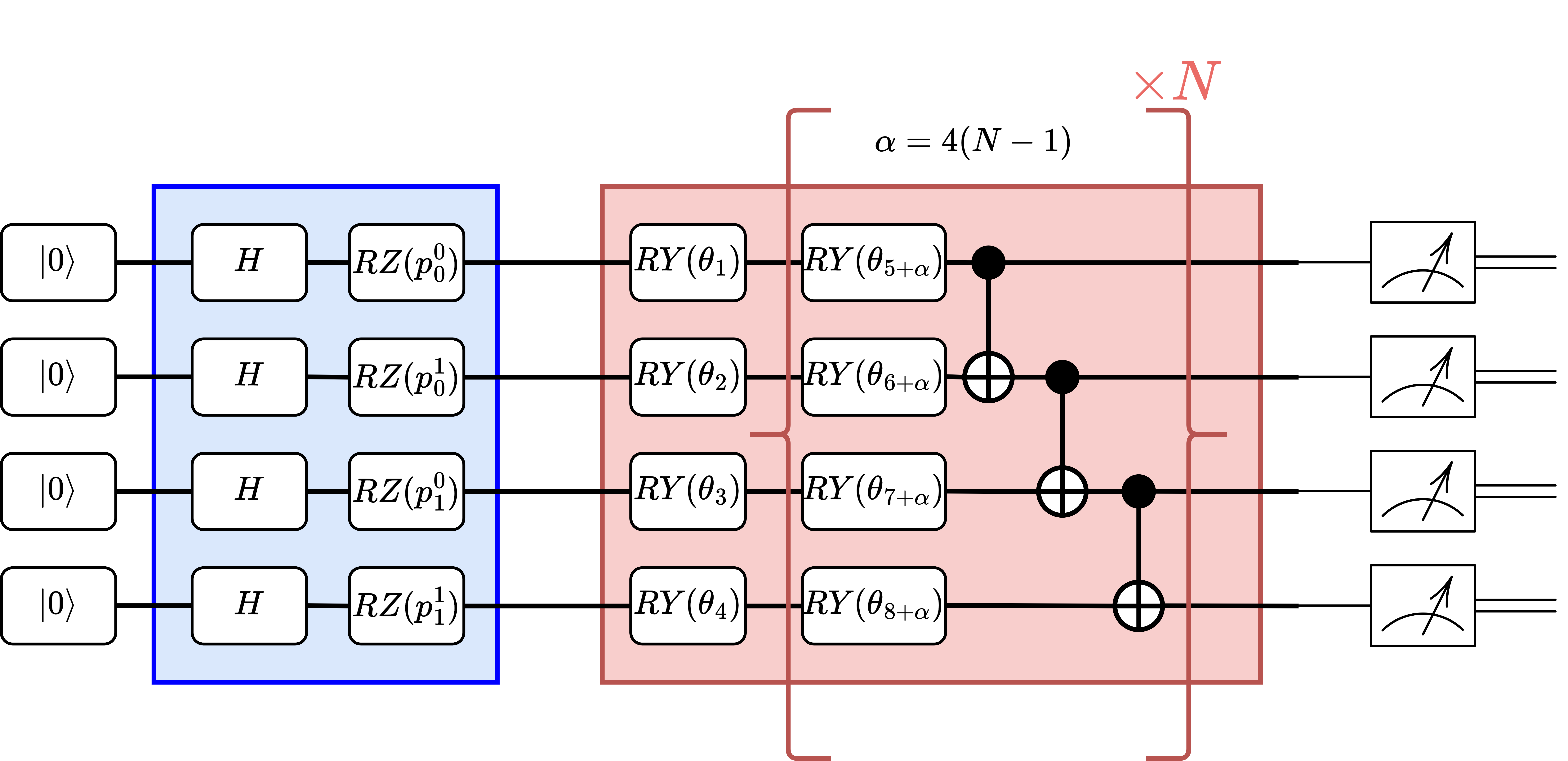}
    \caption{An example of the baseline QNN model that uses the $Z$ feature map and a hardware efficient ansatz. The data-encoding part of the circuit in the blue box encodes two 2D vectors using the $Z$ feature map. The ansatz in the red brackets is repeated $N$ times.}
    \label{fig:baseline}
\end{figure*}

In this subsection, we introduce the basic QNN~\cite{QNN} as a baseline model to deal with point cloud datasets. For a typical point cloud dataset, each input data consists of a series of vector points,  the value of which indicates the geometrical positions of objects in space. In this paper, we denote a vector in the point cloud by $\vec{p}$. For example, a two-dimensional point cloud dataset contains a collection of $x$ and $y$ axis coordinates of data points in the form of: $\{\vec{p}_{i}\}=\{(p_{i}^{0},p_{i}^{1})\}$, where the superscript denotes the coordinate index and the subscript denotes the index of the data point. To use these inputs for a QNN, we first flatten the input into a one-dimensional array, $\{p_{0}^{0},p_{0}^{1},p_{1}^{0},p_{1}^{1},...,p_{n}^{0},p_{n}^{1}\}$, and then encode these values in qubits using $Z$ feature maps. A $Z$ feature map consists of single qubit $Z$-rotations with angles corresponding to the values of the input array. If the input array is normalized to have values within $[0,2\pi]$, the $Z$ feature map will then offer a unique encoding that maps the input array onto the quantum state of qubits:~$\lvert \psi(\vec{p})\rangle$. For a dataset consisting of $n$ data points in the $d$ dimensional space, it will require $N_{q} = n\cdot d$ number of qubits to fully encode the input data. After the input data has been encoded onto the qubits, a QNN will then use multiple layers of hardware efficient ansatz, denoted by $U({\vec{\theta}})$, consisting of single qubit rotations and CNOT gates, as shown in Figure~\ref{fig:baseline}.

\subsubsection{QNN with rotational symmetry}
\label{ssec:rotation}
It is obvious that the baseline QNN circuit given in Figure \ref{fig:baseline} is not rotationally invariant.  To construct a circuit with the exact rotational symmetry, we first introduce a theorem that provides some insights into this problem. From Weyl's work~\cite{weyl1946classical} and its application in modern ML architecture~\cite{villar2023scalars, PELICAN}, we note that for any rotationally invariant function with vector inputs, $I(p_{1},...,p_{N})$, there always exists an equivalent rotationally invariant function $F(\{p_{i}\cdot p_{j}\})$ with pair-wise inner products as inputs:

\begin{equation}
\label{eq:weyl}
    I(p_{1},...,p_{N}) = F(\{p_{i}\cdot p_{j}\}_{i,j}).
\end{equation}

This suggests that the pair-wise inner products provide enough information to build any rotationally invariant function that could be built with the vector inputs. Although such a change in the inputs' format might seem trivial at first, a closer examination will reveal that this approach offers a significantly easier way of encoding the rotational symmetry. On the left-hand side of Equation \eqref{eq:weyl}, the function, $I$, has to be specifically chosen to have rotational invariance, which, in the context of NN, often involves layers of convolutions. However, the right-hand side of Equation \eqref{eq:weyl} is a function that has the rotational symmetry guaranteed at the input level since all inner products are rotational invariants. This provides great freedom in the construction of $F(\{p_{i}\cdot p_{j}\})$, the architecture of which has no constraint as far as the rotational symmetry is concerned. This is particularly beneficial in QNNs since it allows for the use of hardware-efficient ansatz in the construction of a rotationally invariant circuit. In classical machine learning, this approach was used to build the PELICAN network, which can identify top quarks at the state-of-the-art $94.25\%$ accuracy~\cite{PELICAN}.

As an example, we consider the input of four spatial points in arbitrary dimension space. The circuit in Figure \ref{fig:symmetry_circuit}  computes the pair-wise inner products, $p_{i}\cdot p_{j}$, and uses them as the inputs in the feature map encoding shown in the blue box.

\subsubsection{QNN with rotational and permutation symmetries}
\label{ssec:permutation}

\begin{figure*}
    \centering
    \includegraphics[width=1.0\textwidth]{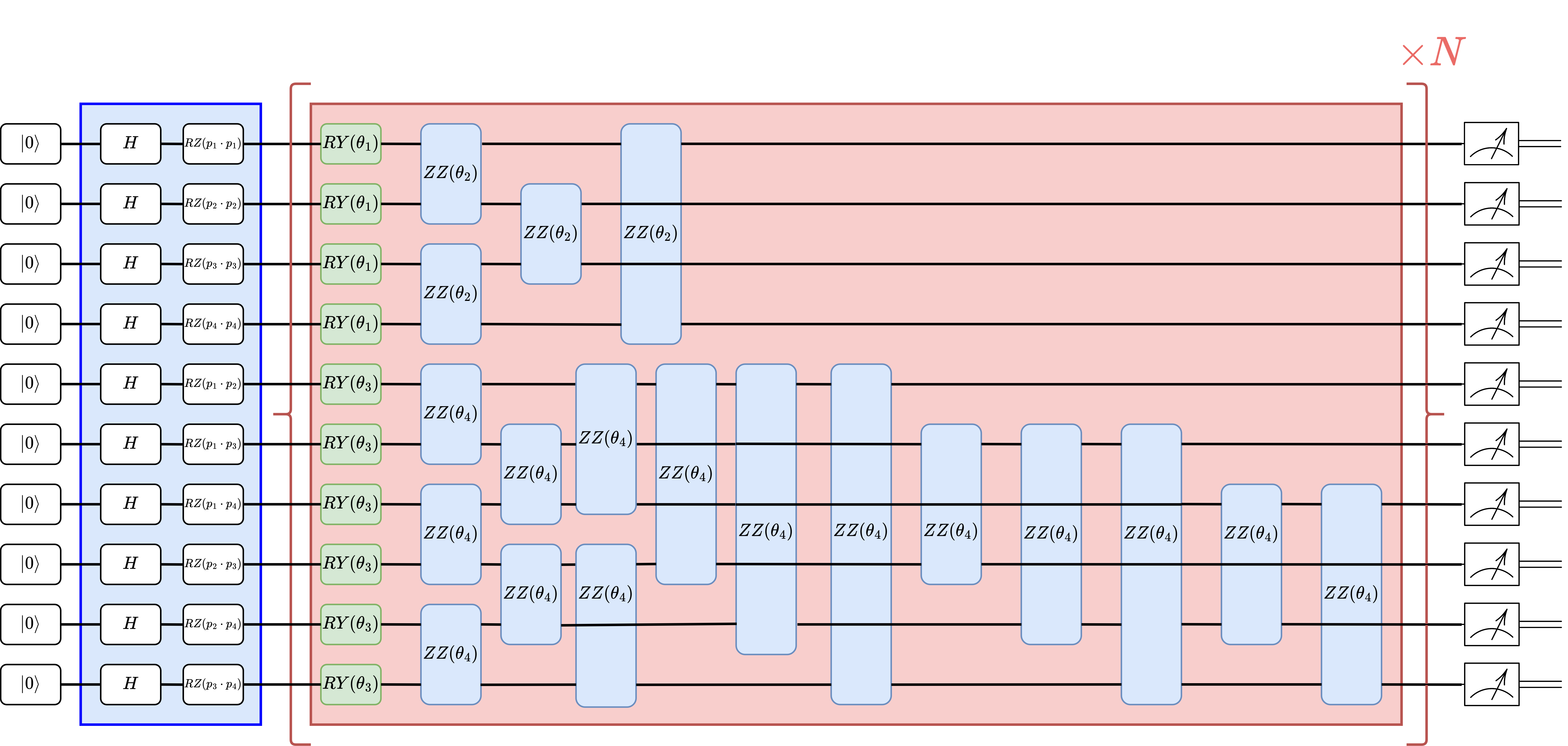}
    \caption{An example of a QNN setup with both rotational and permutation symmetry of four input vectors. The rotational symmetry is enforced by having pair-wise inner products as inputs, which requires $4\cdot(4+1)/2=10$ qubits in the encoding process: $\{p_{1}\cdot p_{1}\}, ..., \{p_{3}\cdot p_{4}\}$. The ansatz is decomposed into two blocks as discussed in Section \ref{ssec:permutation}.  Each twirled operator is required to have the same parameter by the twirling method to achieve equivariance.  Each $ZZ$ gate takes exactly two inputs. If a horizontal line goes through a $ZZ$ gate, it implies that the corresponding qubit is not one of the $ZZ$ gate's inputs.}
    \label{fig:symmetry_circuit}
\end{figure*}

For point cloud datasets, there often exists a permutation symmetry, in which the label is invariant under permutations of the input data points. Suppose that the data points are invariant under the symmetry induced by a group $\mathcal{S}$, which could be taken to be the permutation group in this section.
Let $V[s]$ be a representation of the symmetry group $\mathcal{S}$ on the data points space $\{p_i\}$.
The goal here is to construct a QNN circuit which satisfies
\begin{equation}
    \hat{y}(V[s] \cdot p; \theta)
    =
    \hat{y}(p;\theta)\,,
\end{equation}
where we denote $p=\{p_i\}$, and $V[s] \cdot p$ is the symmetry group elements acting on the data points space. The QNN with this condition is called an equivariant QNN.

First, the action of symmetry group on the encoding unitary can be expressed using the so-called induced representation:
\begin{equation}
    U_{\text{E}}(V[s] \cdot p)
    =
    V^{\text{(ind)}}[s] U_{\text{E}}(p) V^{\text{(ind)}\dag}[s]\,.
\end{equation}
Then, our task is to symmetrize QNN unitary $U_{\text{QNN}}$ using $V^{\text{(ind)}}[s]$. Suppose that our QNN is composed of the products of $G$ from group $\mathcal{G}$: $R_G(\theta)=\exp(-iG \theta)$. Then the symmetrization is equivalent to 
\begin{align}\label{eq:symmetrization}
    [R_G(\theta), V^{\text{(ind)}}[s]]&=0,\quad\text{for all} \quad\theta\in[0,2\pi], G\in \mathcal{G}, s\in\mathcal{S}\,,
    \\
    \Leftrightarrow
    [G, V^{\text{(ind)}}[s]]&=0,\quad\text{for all} \quad G\in \mathcal{G}, s\in\mathcal{S}\,.
\end{align}

A common approach to symmetrize generators $G\in\mathcal{G}$ is the use of the twirling formula~\cite{QNN_permutation,nguyen2022theory, Meyer_2023}, in which  $\mathcal{T}[G]$ is equivariant to all elements in group $\mathcal{S}$.

\begin{equation}\label{eq:twirl}
    \mathcal{T}[G]
    =\frac{1}{|{\mathcal{S}}|}
    \sum_{s\in\mathcal{S}}
    V^{\text{(ind)}}[s]GV^{\text{(ind)}\dag}[s]
\end{equation}

Then the resulting QNN output $\hat{y}$ becomes equivariant, if the observable commutes with the induced unitary operators. For a one-index input vector, the twirl operator simply requires the operator at hand,~$G$,  to be looped through all permutations of the input vector elements to achieve equivariance. In our network with pair-wise inner products between data points described in Section~\ref{ssec:rotation}, we have two-indexed objects, $p_{i}\cdot p_{j}$, as inputs. For convenience, we will from now on denote the two-indexed object with the notation: $q_{ij} = p_{i}\cdot p_{j}$.  The transformation of the pair-wise inner products with respect to the permutation of the datapoints is not the same as those considered in References~\cite{QNN_permutation,nguyen2022theory, Meyer_2023}. To obtain some intuitions about the permutation of the two-indexed objects, we first decompose the vector $\vec{Q} = \{ q_{ij}\}$ into two separate blocks of elements: the self-interaction terms, $\vec{Q}_{\text{self}}=( q_{11}, ..., q_{ii})$, and the pair-wise interaction terms, $\vec{Q}_{\text{pair}}= ( q_{12}, ..., q_{ij})$ for $i \neq j$. Under the $S_{n}$ permutation of point cloud data points, we observe that both blocks transform into themselves. As a result, these two blocks of terms are respectively transforming like $S_{n}$-permutated one-index vectors. It is then easy to construct permutation-equivariant ansatz in both blocks by applying the twirling formula separately. In Appendix~\ref{appendix:n3_perm}, we give the full detail of constructing such an ansatz for the case of three input vectors.

The circuit in Figure \ref{fig:symmetry_circuit} is an example of permutation and rotationally invariant QNN with four vectors as inputs. We see that the inputs are divided into two blocks: the first four elements are the self-interactions and the remaining terms are the pair-wise interactions. The ansatz is built by twirling the Pauli-$Y$ rotations and the entangling $ZZ$ rotations in both blocks.

\subsection{Loss function}
\label{ssec:nonlinear}

The output of the QNN, $\hat{y}(p_{i};\theta)$, is compared to the truth label of the events to perform classification.  The loss function is taken to be the standard Mean Squared Error (MSE) defined as:
$\mathcal{L}(\theta)= \frac{1}{4n}\sum_{i=0}^{n} (\hat{y}(p_{i};\theta) -y_{i})^{2}$, where $n$ is the size of the training set and the extra factor of $4$ is used to normalize the MSE into the range of $[0,1]$ given that the output is between $-1$ and $1$, and the labels are given by $y \in \{-1,1\}$. 
In practice, many real-world problems require nonlinear activation of the input data. In such cases, we introduce an ad-hoc activation function at the end of the QNN as follows. 

\begin{equation}
\begin{split}
\label{eq:nonlinearLoss}
    \mathcal{L'}(\theta,b)=\sum_{i=0}^{n}(-|\hat{y}(p_{i};\theta)-b|-y_{i})^{2},
\end{split}
\end{equation}
where $b$ is an offset parameter to be optimized by the learning process.  The function has a discontinuity at $\hat{y}(p_{i};\theta)=b$ by introducing the absolute difference between $\hat{y}(p_{i};\theta)$ and $b$.  In this paper, we consider the task of applying  QNN to identify signal data, labeled by  $y=1$, out of datasets containing a mixture of signal and background data, in which the background data is labeled by $y = -1$. Intuitively, the activation function in Equation~\eqref{eq:nonlinearLoss} will be handy if the QNN output of the signal data is concentrated at a particular value while background data has QNN outputs on both sides. In such cases, the function $-|\hat{y}(p_{i};\theta)-b|$ will learn to spot such signal concentrations. Since the offset parameter is explicitly defined in the cost function, its derivative is well-defined for all values except at the exact discontinuity. In each iteration, we update $b$, similar to other parameters, through either gradient descent or COBYLA.

\section{Results}
\label{sec:results}
In this section, we provide the experimental results of the implementation of the symmetric QNNs. In Section~\ref{ssec:2Drot}, we create datasets with 2D rotations of certain geometrical figures and test the performance of QNN with symmetry against that of the baseline QNN. In Section~\ref{ssec:decay}, we consider the particle decay classification, which involves the $SO(1,3)$ Lorentz symmetry. In Section~\ref{ssec:BP}, we investigate the trainability of our models by numerically evaluating the variance of the first-order gradients upon random initializations.
All numerical results presented here are obtained by the classical quantum circuit simulator implemented via Qiskit~\cite{Qiskit}.

\subsection{Classification of 2D rotations}
\label{ssec:2Drot}

\begin{figure}[H]
     \centering
     \begin{subfigure}[b]{0.22\textwidth}
         \centering
         \includegraphics[width=\textwidth]{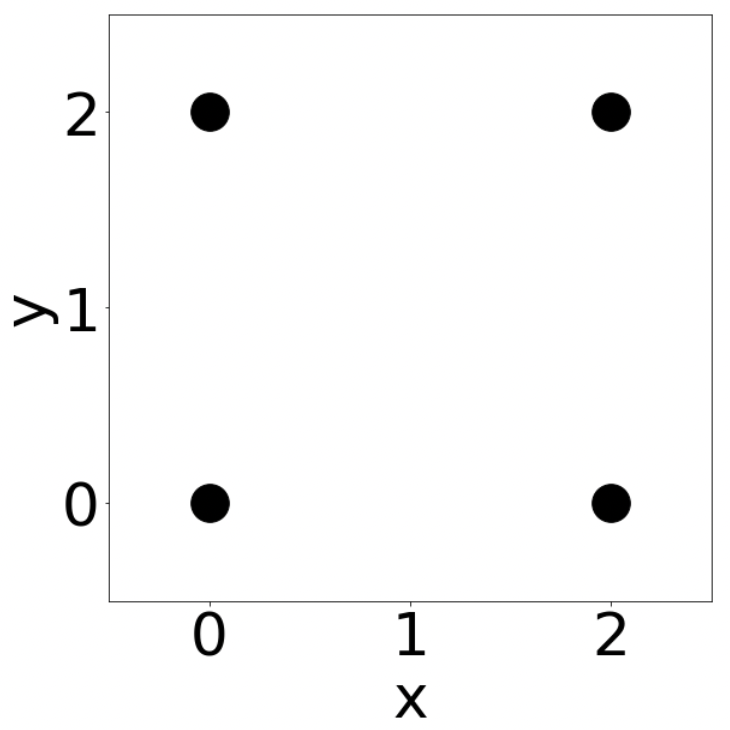}
         \caption{Square template}
         \label{fig:Square_template}
     \end{subfigure}
     \hfill
     \begin{subfigure}[b]{0.22\textwidth}
         \centering
         \includegraphics[width=\textwidth]{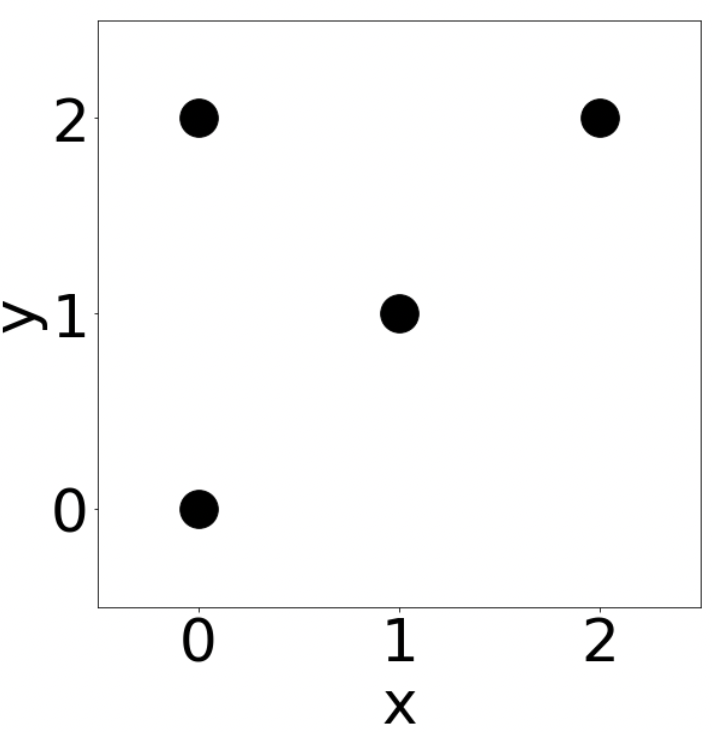}
         \caption{Triangle template}
         \label{fig:Triangle_template}
     \end{subfigure}
     \hfill
     \begin{subfigure}[b]{0.22\textwidth}
         \centering
         \includegraphics[width=\textwidth]{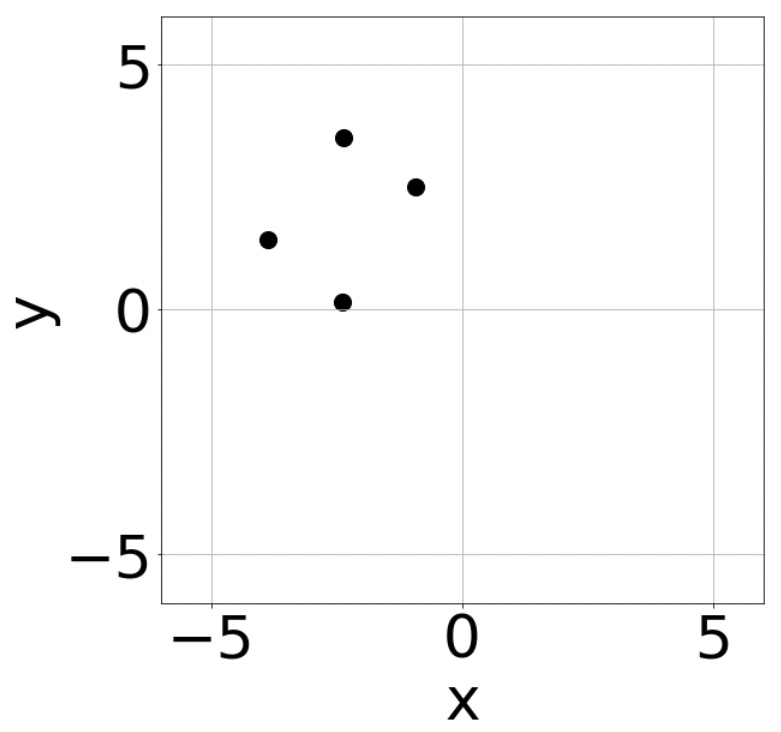}
         \caption{Transformed square.}
         \label{fig:Transformed_square}
     \end{subfigure}
        \hfill
     \begin{subfigure}[b]{0.22\textwidth}
         \centering
         \includegraphics[width=\textwidth]{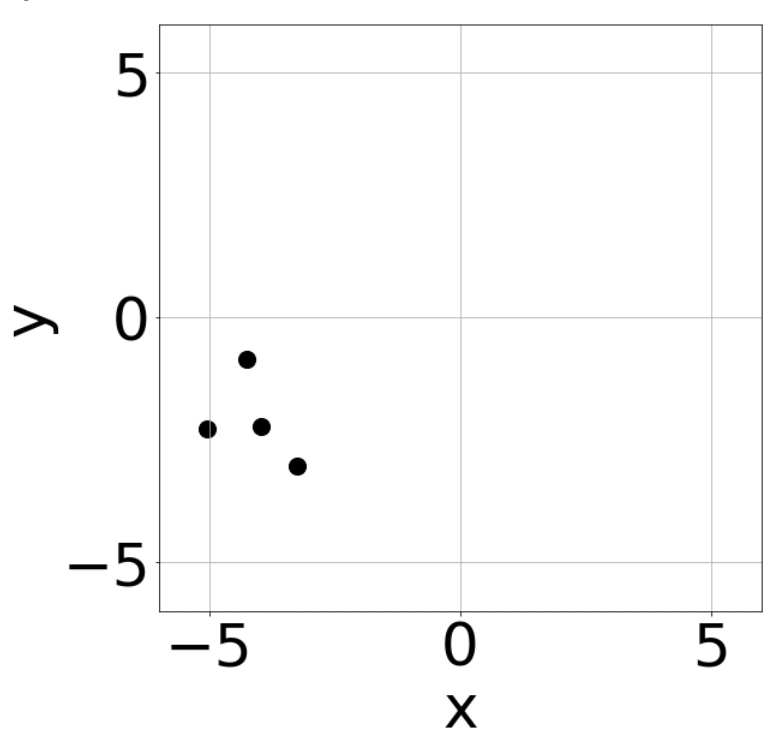}
         \caption{Transformed triangle.}
         \label{fig:Transformed_triangle}
     \end{subfigure}
        \caption{Templates and transformed images of the 2D dataset containing squares and triangles. The templates of the two geometrical objects are shown in the first two subfigures at the top. The two images at the bottom show two examples of transformed images, in which the template went through spatial translation, re-sizing, rotations, and smearings.}
        \label{fig:three graphs}
\end{figure}

To test the performance of our new architecture, we first create a dataset with rotated images of two geometrical objects: the square and the triangle. The templates of these two objects are shown in Figure \ref{fig:Square_template} and Figure \ref{fig:Triangle_template}, respectively. The two templates consist of four data points, three of which are placed in the same locations. To further impose hurdles on the learning process, we consider various transformations to be applied to these templates. The transformations are summarized in the following list:

\begin{figure*}
     \centering
     \begin{subfigure}[b]{0.45\textwidth}
         \centering
         \includegraphics[width=\textwidth]{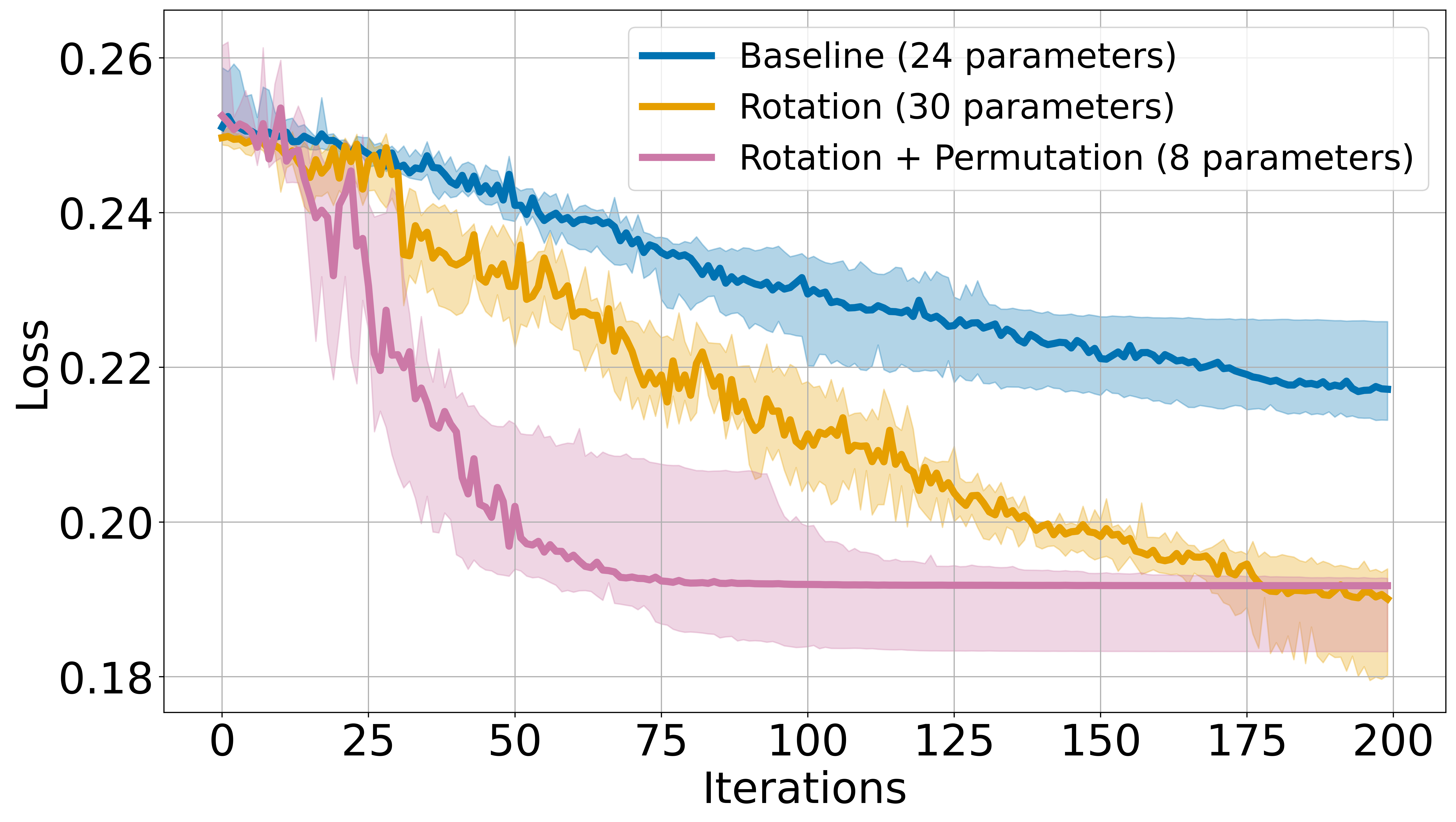}
         \caption{Loss function vs iterations.}
         \label{fig:loss_2d}
     \end{subfigure}
     \hfill
     \begin{subfigure}[b]{0.45\textwidth}
         \centering
         \includegraphics[width=\textwidth]{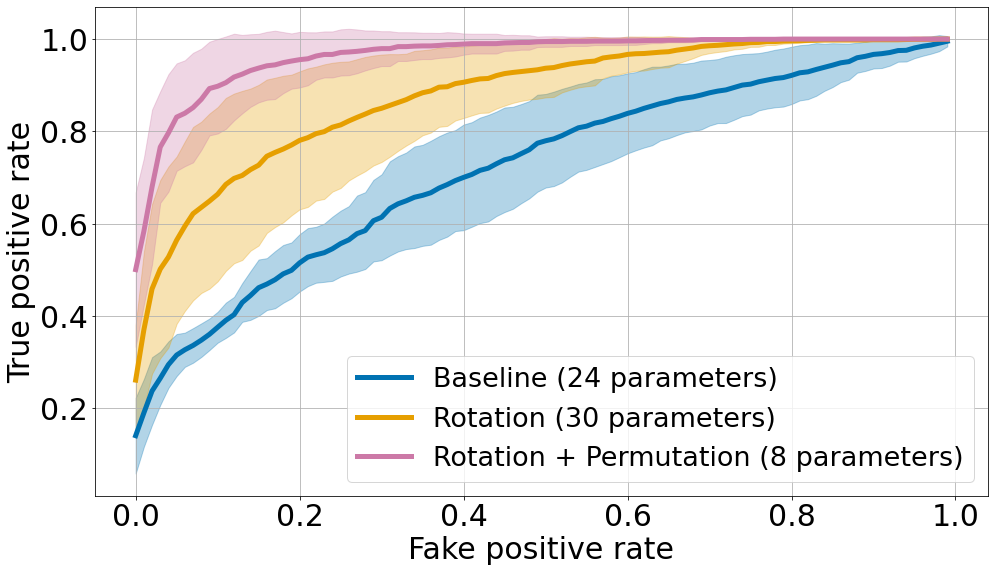}
         \caption{ROC curve.}
         \label{fig:ROC_2d}
     \end{subfigure}

        \caption{The Loss function during the training process and the ROC curve for various QNN models in the 2D image classification task. The loss values shown in the figure are the median values of all ten initializations, whereas the error bands are computed at $25\%$ and $75\%$ quantiles. The ROC curves are obtained by averaging the training results and the error bands are obtained by taking the standard deviations.}
        \label{fig:loss_ROC_2d}
\end{figure*}

\begin{table*}
\centering
\begin{tabular}{ |c|c|c|c| } 
\hline
Methodology & Baseline & Rotation & Rotation + Permutation \\
\hline
$\#$ parameters & 24 & 30  & 8 \\ 
\hline
$\#$ qubits & 8 & 10 & 10\\
\hline
Depth & 2 & 2 & 12\\
\hline
AUC & $0.720 \pm 0.060$ & $0.880 \pm 0.070$ & $0.966 \pm 0.030$\\
\hline
\end{tabular}
\caption{Specifications and performances of the QNNs used in the 2D image classification task.}
\label{tab:2D}
\end{table*}

\begin{itemize}
    \item \textit{Translation}: The data points in the templates are translated collectively as a group in both the~$x$ and~$y$-axis. The maximum translational distance is set at 5.
    \item \textit{Re-sizing}: The size of the template is varied by a factor randomly drawn between 0.5 and 5.
    \item \textit{Rotation}: The template is rotated around its center by an arbitrary angle.
    \item \textit{Shuffling}: The order of data points in the dataset is subject to random shuffling such that there is no preferred order in the four data points.
    \item \textit{Smearing}:  Each data point in the template is smeared such that it is moved around its nominal position by a random amount drawn between -0.5 and 0.5. The smearing is performed independently for each data point.
\end{itemize}

Two examples of the transformed images are shown in Figure \ref{fig:Transformed_square} and Figure \ref{fig:Transformed_triangle}. Overall, we created $1,600$ different images, each of which is a template undergone all five random transformations listed above. In this dataset, $1,200$ images are used as training data while the other $400$ are used as testing data to verify performances.

\begin{figure*}
     \centering
     \begin{subfigure}[b]{0.45\textwidth}
         \centering
         \includegraphics[width=\textwidth]{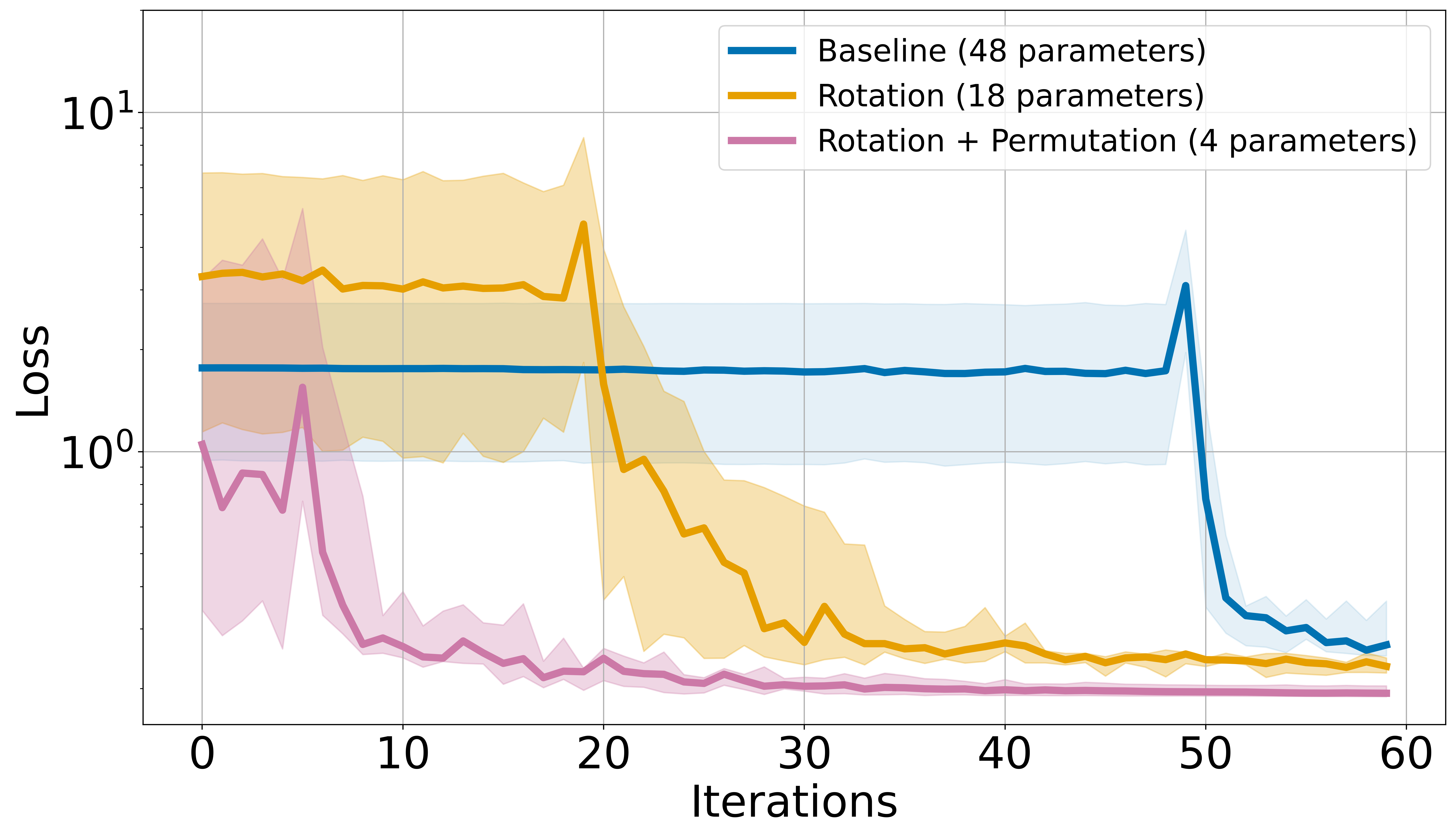}
         \caption{Loss function vs iterations.}
         \label{fig:loss_decay}
     \end{subfigure}
     \hfill
     \begin{subfigure}[b]{0.45\textwidth}
         \centering
         \includegraphics[width=\textwidth]{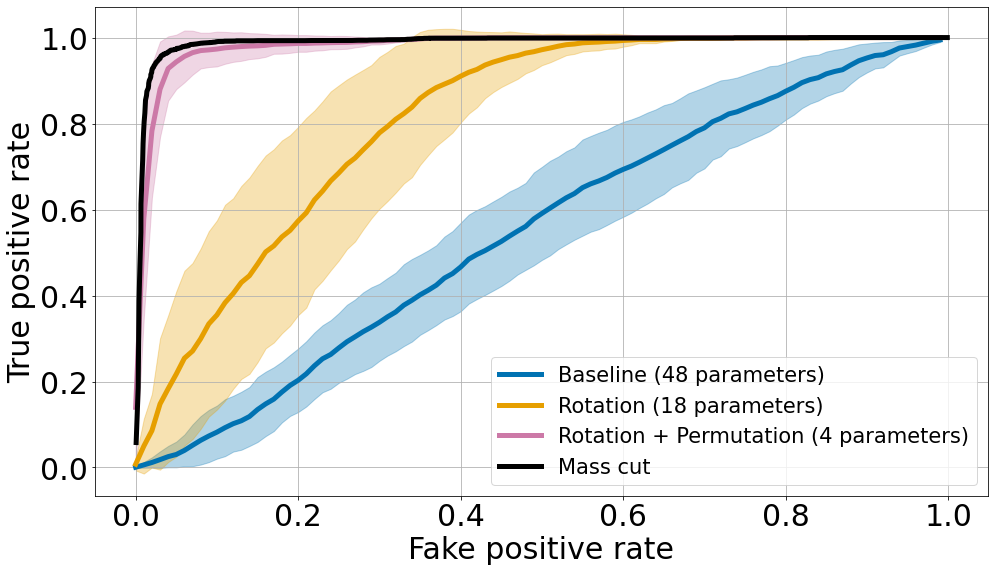}
         \caption{ROC curve.}
         \label{fig:ROC_decay}
     \end{subfigure}

        \caption{The Loss function during the training process and the ROC curve for various QNN models in the particle decay identification task. The loss values shown in the figure are the median values of all ten initializations, whereas the error bands are computed at $25\%$ and $75\%$ quantiles. The ROC curves are obtained by averaging the training results and the error bands are obtained by taking the standard deviations. In the ROC curve, the QNN results are compared to the performance of the mass cut, which is regarded as the theoretical upper limit of performance.}
        \label{fig:loss_ROC_decay}
\end{figure*}
 
\begin{table*}
\centering
\begin{tabular}{ |c|c|c|c|c| } 
\hline
Methodology & Baseline  & Rotation & Rotation + Permutation & Mass cut \\
\hline
$\#$ parameters & 48 & 18  & 4 & N/A \\ 
\hline
$\#$ qubits & 16 & 6 & 6 & N/A\\
\hline
Depth  & 2 & 2 & 12 &  N/A\\
\hline
AUC & $0.544 \pm 0.060$ & $0.813 \pm 0.076$ & $0.982 \pm 0.011$  & $0.988 \pm 0.001$\\
\hline
\end{tabular}
\caption{Specifications and performances of the QNNs used in the particle decay classification task.}
\label{tab:decay}
\end{table*}

	We consider three models to tackle this 2D image classification problem: the baseline model with no symmetry, the QNN model with rotational symmetry, and the QNN model with full permutation and rotational symmetries. The baseline model is the same as the one described in Section \ref{ssec:QNN}. The input of the baseline model uses the flattened array containing the $x$ and $y$ coordinates of the data points in the image. The baseline model employs $N_{\text{base}} = 2$ layers of hardware efficient ansatz with 24 parameters in total. We also consider a model that respects the rotational symmetry but not the permutation symmetry. The rotationally symmetric model uses pair-wise inner products between the four data points as inputs, giving $4\cdot(4+1)/2=10$ qubits in total. This model also uses $N_{\text{rot}} = 2$ layers of hardware efficient ansatz, which gives 30 parameters in total. Finally, we construct the model with the full rotational and permutation symmetries according to the recipe shown in Figure \ref{fig:symmetry_circuit}. It is clear that, due to the symmetry constraints, the ansatz in the fully symmetric model contains a much smaller number of parameters. In practice, the fully symmetric model has $N_{\text{full}} = 2$ layers of the ansatz with 8 parameters in total. However, this fully symmetric model appears to be much more effective in the classification task when compared to the baseline one. To obtain numerical results, we train each model ten times, each with a different initialization. The training results of all three models are shown in Figure \ref{fig:loss_2d}, where the regression loss is plotted against the number of iterations in training. The median values of all initializations are shown in the figure with error bands computed at the $25\%$ and $75\%$ quantiles. The regression loss is taken to be the MSE loss, as defined in Section~\ref{ssec:nonlinear}, across the whole dataset. The fully symmetric model is observed to have converged much faster than the two alternative models. In terms of performance, we quantitatively measure the classification performance using the Receiver Operating Characteristic (ROC) curve shown in Figure \ref{fig:ROC_2d}. The performances of the trained models are quantitatively measured by the Area Under the Curve (AUC) of their ROC curves, which are listed in Table \ref{tab:2D}~\footnote{We also perform simulation for the baseline and rotation equivariant models with the number of layers and the number of iterations increased but the fully symmetric model presented here still gives the best performance. The detail is given in Appendix~\ref{appendix:alternative-settings}.}. The circuit depth shown in Table \ref{tab:2D}, obtained directly via \textit{Qiskit}~\cite{Qiskit}, is a measure of the circuit's effective number of layers during parallel execution. The QNN model with full rotation and permutation symmetries has the best performance with an AUC of $0.966 \pm 0.030$.

\subsection{Classification of high energy particle decay with ${SO}(1,3)$ Lorentz symmetry}
\label{ssec:decay}

Another task we have considered is the classification of particle decays. We use a simulated Monte Carlo dataset that corresponds to the running conditions at the large hadron collider, where protons collide at high energies to produce various high-energy particles that subsequently decay~\cite{opendata}. In this task, we use the Higgs boson decay as the signal process, where a Higgs boson decays into four final state leptons through the $ZZ^{*}$ channel: $H\xrightarrow[]{}ZZ^{*}\xrightarrow[]{}\ell^{\pm}\ell^{\mp}\ell^{\pm}\ell^{\mp}$. The background consists of four lepton events from various Standard Model processes excluding Higgs boson production.  A total number of $6,000$ particle decay events are used in the training, while $2,000$ events are used for testing performances.  For the baseline model, we use the four momenta of the final state leptons as inputs. To enforce rotational symmetry, we acknowledge that the ${SO}(1,3)$ Lorentz symmetry also satisfies the theorem given in Equation~\eqref{eq:weyl}~\cite{weyl1946classical,villar2023scalars, PELICAN} and thus use the pair-wise inner products between final state particles as inputs. In this task, the invariant masses of leptons are negligible compared to the energy scale of the decay, which is at the scale of the Higgs mass. As a result, it is convenient to drop the self-interacting terms in the first block of Figure~\ref{fig:symmetry_circuit}. In practice, this implementation does not show any loss of performance when compared to the model with the self-interactions.  To further increase the performance of the QNN models, we implement the activation function in Equation~\eqref{eq:nonlinearLoss}  on all QNN models for the particle decay classification problem.

The training loss is shown in Figure~\ref{fig:loss_decay}, in which we observe that the fully symmetric model again converges much faster. In the ROC curve shown in Figure~\ref{fig:ROC_decay}, we also include the traditional classification method of window cuts on the four-body invariant mass. Since this traditional method directly yields the Higgs mass and cuts around it, it could be viewed as the most optimal method given the available information.  The AUC of all models is shown in Table~\ref{tab:decay}, in which the fully symmetric model has achieved similar performance as that of the mass cut method.

\begin{figure*}
     \centering
     \begin{subfigure}[b]{0.3\textwidth}
         \centering
         \includegraphics[width=\textwidth]{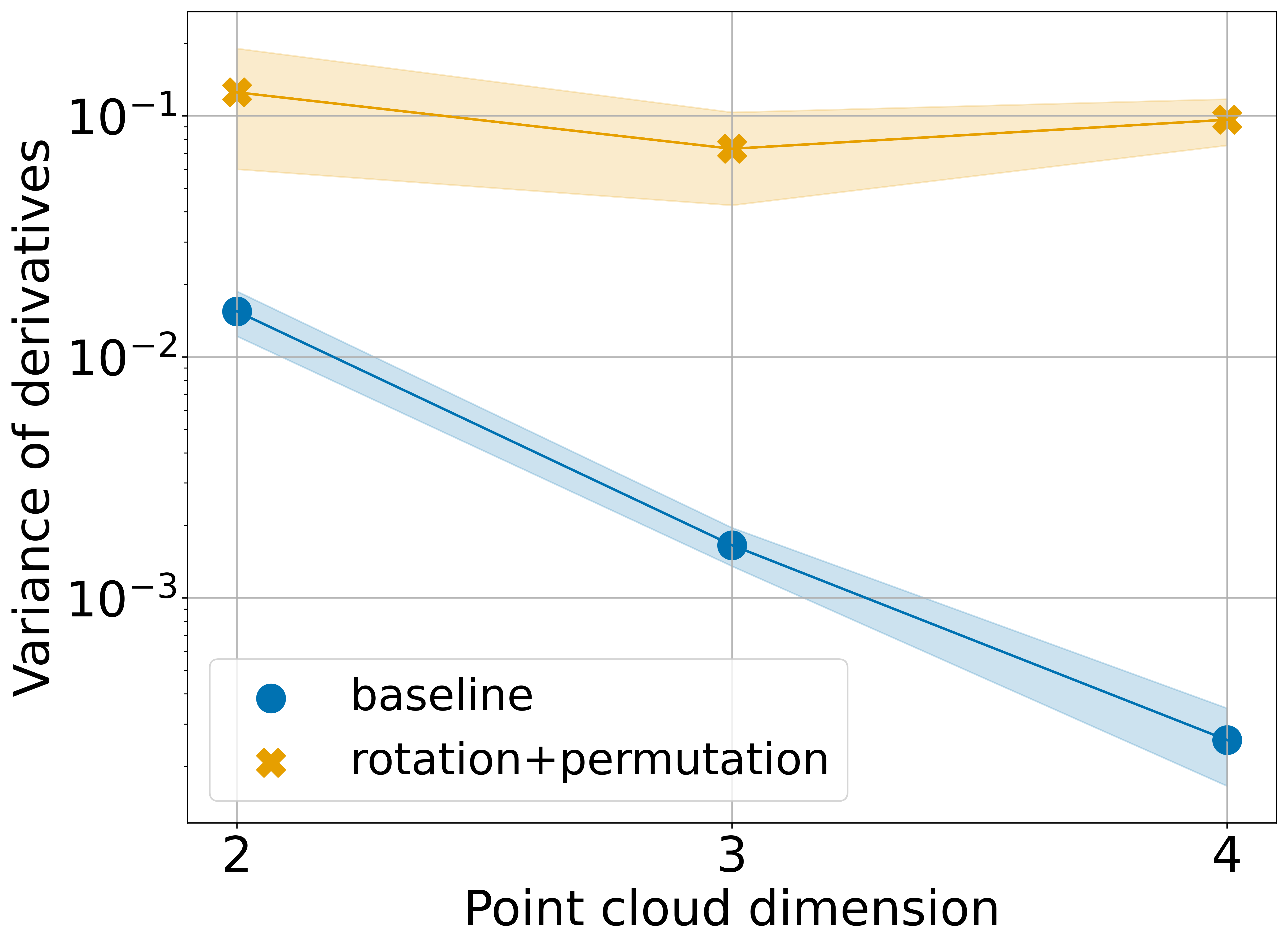}
         \caption{Derivative variance versus input dimensions.}
         \label{fig:var_vs_dimension}
     \end{subfigure}
     \hfill
     \begin{subfigure}[b]{0.3\textwidth}
         \centering
         \includegraphics[width=\textwidth]{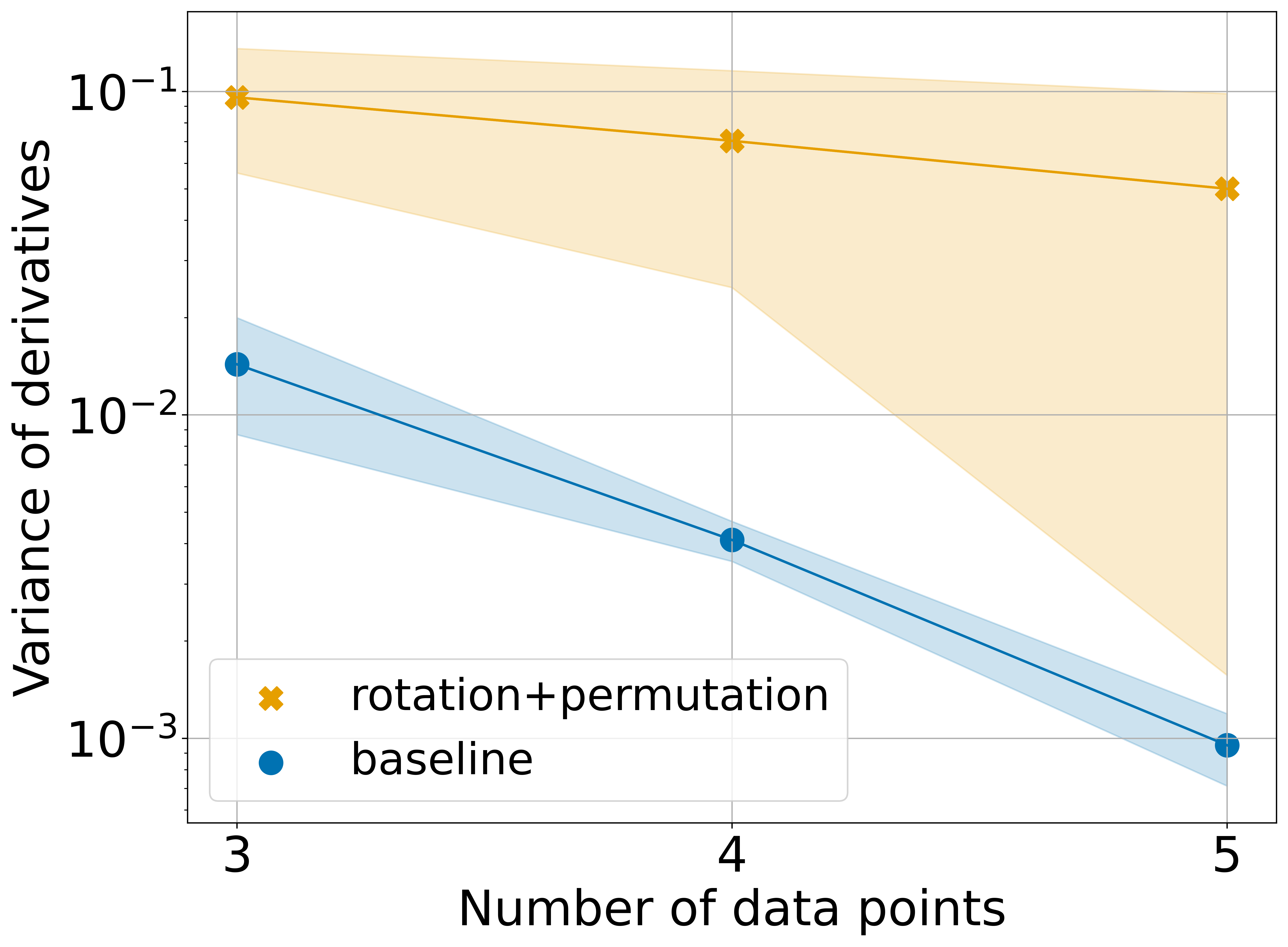}
         \caption{Derivative variance versus number of input data points.}
         \label{fig:var_vs_nParticles}
     \end{subfigure}
          \hfill
     \begin{subfigure}[b]{0.3\textwidth}
         \centering
         \includegraphics[width=\textwidth]{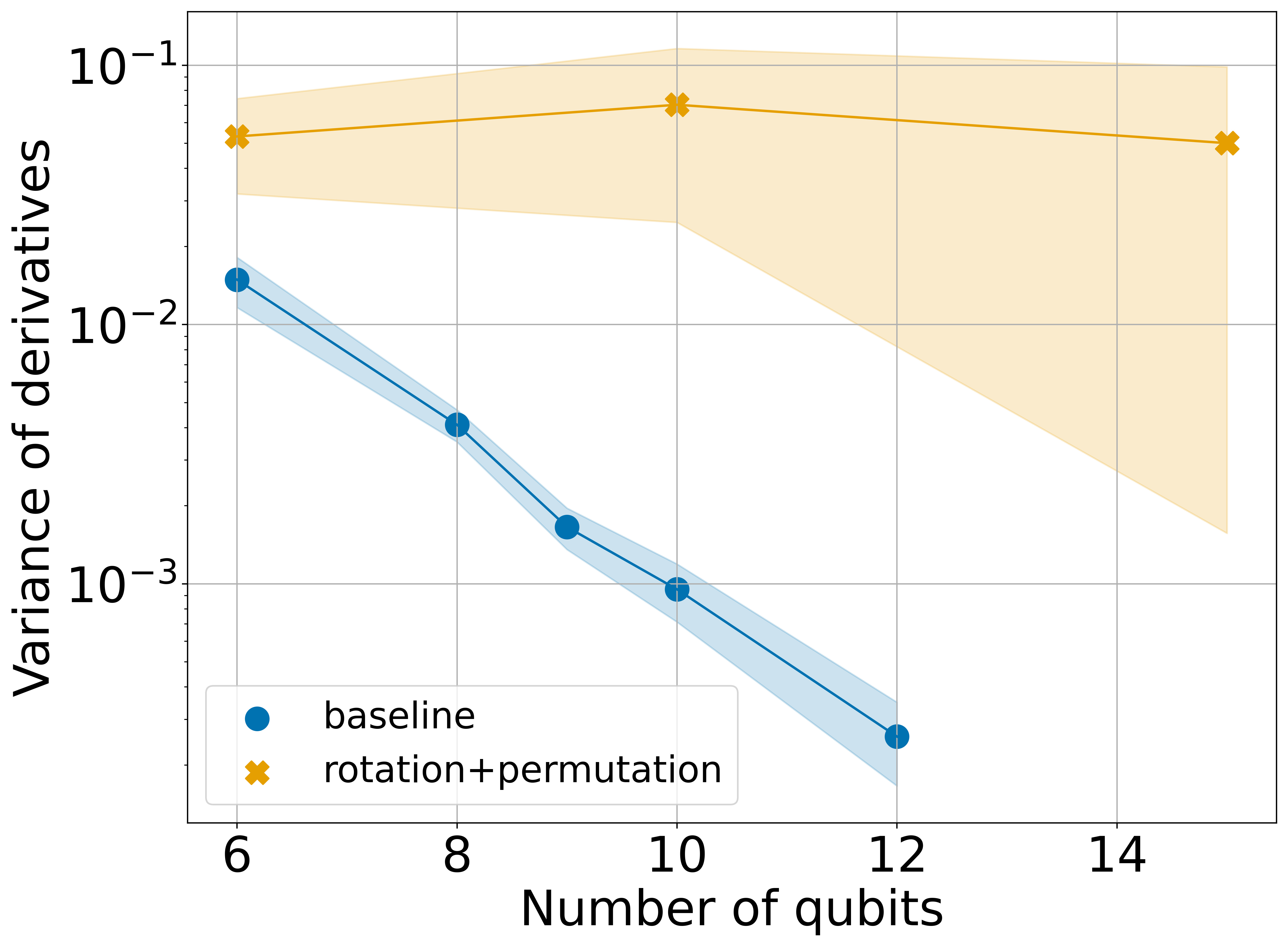}
         \caption{Derivative variance versus number of qubits.}
         \label{fig:var_vs_nqubits}
     \end{subfigure}

        \caption{The variance of first-order derivatives of the QNN output. The dimension and number of data points in the input data are varied to observe their impact on the gradients. The variance is computed by sampling $100$ randomly initiated parameters. The third plot shows the number of qubits used in the ansatz on the x-axis. }
        \label{fig:variance}
\end{figure*}

\subsection{Barren plateau}
\label{ssec:BP}

In previous sections, the QNN with the full symmetries has been observed to converge much faster during the training procedures. This can be further investigated by checking the variance of the first-order gradients, which is exponentially suppressed for particular models, such as circuits with large depths, high expressibility, and high entanglement (see e.g.~\cite{McClean_2018_BP,expressiveness_Holmes_2022,costFunctionBP_2021, PhysRevResearch.3.033090, PhysRevLett.128.180505, PRXQuantum.2.040316}). We take the circuits used in Section \ref{ssec:2Drot} and vary the dimensionality and number of input data points to observe their impact on the trainability of our models.

First, we fix the number of input data points to three and vary the dimension of the point cloud input between two and four. The first-order derivatives are sampled based on~$100$ randomly initiated sets of parameters. Since the baseline model requires more qubits as the dimension increases, we can easily observe the barren plateau effect in Figure~\ref{fig:var_vs_dimension}, where the derivative variance of the baseline model decreases exponentially. However, since our symmetric QNN model uses inner products as inputs, the increase in dimensions has practically no impact on the gradients. We then proceed to fix the dimension of the input at two and vary the number of data points between three and five. Again, as shown in Figure~\ref{fig:var_vs_nParticles}, the baseline model exhibits clear signs of a barren plateau due to its expressive nature. In the case of the symmetric model, it does not appear to be expressive enough for the gradient to vanish since we only used two layers. The exploitation of symmetries has allowed us to use a much shallower circuit ansatz with fewer parameters that had proven to be much more trainable. Finally, we combine the first two plots into Figure~\ref{fig:var_vs_nqubits} to show that the baseline model observes a clear barren plateau phenomenon where the variance of derivatives decreases exponentially as the number of qubits increases. In our test case, the symmetrization process significantly reduces the effective Hilbert space of the symmetric model, leading to a consequent reduction in its expressibility. Therefore, it is intuitive that the barren plateau phenomenon, which occurs in expressive circuits satisfying the 2-design constraint~\cite{McClean_2018_BP}, is not observed in our test case due to these reductions. However, this does not guarantee the removal of barren plateau in all symmetric problems.

\section{Discussion and conclusions}
\label{sec:discussion}
In this work, we have demonstrated the performance of the QNN model with rotational and permutation symmetries. The fully symmetric model has shown great potential in the task of classifying point cloud datasets, where the traditional QNN model without symmetries struggles. The new model exploits the fact that point cloud datasets often exhibit rotational and permutation symmetries, which could be encoded into quantum circuits via taking inner products of data vectors and using permutation equivariant ansatz.  

In the classification tasks of 2D images, the fully symmetric QNN model achieved an AUC of ROC curve above $0.966 \pm 0.030$ while having converged typically within a few dozen iterations. The fast converging is likely a result of the lower number of parameters used in the symmetric circuits. Similarly, in the task of the particle decay classification, the fully symmetric model achieved an AUC of $0.982 \pm 0.011$, which is on par with that of the mass cut method. In both tasks, the number of parameters needed in the symmetric circuit to achieve optimal results is proven to be much lower than that in other QNN models. By construction, the symmetric circuit only searches for solutions to the task that is invariant under the symmetry groups. Intuitively, this greatly restricts the search space, in which the optimization is conducted, and thus yields a much faster convergence speed with a lower number of parameters. This observation is in agreement with classical equivariant NN studies using the same principle in Reference~\cite{PELICAN}.

An obvious limitation of our approach is that the input datasets must be point clouds, in which the transformation of global rotational symmetries is captured by taking inner products. For position-dependent local symmetries, this approach is no longer valid. Another important note on this model is that the number of qubits needed grows polynomially with respect to the number of vectors in the input dataset. For a dataset with $n$ vectors in $d$ dimensional space, the traditional method would require $n\cdot d$ number of qubits, whereas our approach would require $n(n+1)/2$ qubits.  Therefore, our approach will need a lot more qubits when $n \gg d$. This feature of our approach encourages us to apply it in problems with $n \ll d$, where the qubit cost is greatly suppressed. Also, in the appendix of Reference~\cite{villar2023scalars}, the authors showed that, for any invariant function to be expressed by the set of inner products, it is sufficient to use only a subset of an approximate size of $n(d+1)$, which is linear with respect to both $n$ and $d$. It is still an open question for us to determine whether there could be other techniques that further improve the scaling of qubits costs.

As a natural extension of our approach, invariants of discrete groups could also be utilized in data encoding to enforce symmetries in the circuit. This offers an alternative for symmetrization that does not require specific treatment of the circuit ansatz, such as the twirling method. However, it is not yet clear whether there exists an equivalent of Weyl's theorem, as in Equation~\ref{eq:weyl}, for all discrete groups to guarantee the completeness of our approach. For some discrete groups, such as discretized rotational groups, we suspect that Weyl's theorem may still hold. However, a precise general statement on this matter will likely require extensive study beyond the scope of this work. We will leave it as a possible future direction of research.

It would be an interesting study to extend this approach to larger datasets with more complex structures. In particular, we would like to point out that an ensemble of quantum state vectors could be viewed as a form of point clouds. In such cases, global unitary transformations acting on the ensemble are equivalent to rotations of state vectors in the Hilbert space. This hints that our approach might be useful in recognizing features of quantum state ensembles that are invariant under unitary transformations.

We would like to point out that estimating the values of a global observable, such as $Z^{\otimes N_{q}}$, will be challenging due to the measurement cost. However, a local observable typically breaks the permutation equivariance in our setup as previously discussed. A candidate of  symmetry-preserving local observables is listed in footnote~\ref{fn:localOB}: $\sum_{n}Z_{n}$. To facilitate the hardware implementation of our approach, it will be an interesting project to investigate how to make use of local observables in a permutation-equivariant fashion.

Another avenue for future research is the choice of loss function. In our approach, we used an ad hoc discontinuous function, as shown in Equation \eqref{eq:nonlinearLoss}. This particular choice of loss function is neither the only option nor proven to be the optimal one. Investigating various forms of loss functions and their relative performances could be a potentially fruitful area of study.

\vspace*{0.5cm}
\acknowledgments

We would like to thank Dr.~Chase Shimmin for the discussions that inspired this project.  This work is supported by the Center of Innovations for Sustainable Quantum AI (JST Grant Number JPMJPF2221).

\appendix
\onecolumngrid
\section{Construction of permutation equivariant model}\label{appendix:n3_perm}
For $n = 3$, we have:
\begin{equation}
Q_{\text{pair}}
=
\begin{pmatrix}
q_{12}
\\
q_{13}
\\
q_{23}
\end{pmatrix}\,,
\quad
Q_{\text{self}}
=
\begin{pmatrix}
q_{11}
\\
q_{22}
\\
q_{33}
\end{pmatrix}\,.
\end{equation}
The representation matrices act on these vectors independently, in other words, the representation is reducible: $V= V_{\text{pair}}\oplus V_{\text{self}}$.
\subsection{Pair vector}
The corresponding representation of ${S}_{3}$ acting on the ``pair'' vector is obtained as
\begin{align}
V_{\text{pair}}[\mathbb{I}_{{S}_{3}}] & = 
\begin{pmatrix}
 1 & 0 & 0
 \\
 0 & 1 & 0
 \\
 0 & 0 & 1
\end{pmatrix} \,,
&
V_{\text{pair}}[\sigma_{(12)}] & = 
\begin{pmatrix}
 1 & 0 & 0
 \\
0 & 0 & 1
\\
0 & 1 & 0
\end{pmatrix} \,,
\\
V_{\text{pair}}[\sigma_{(23)}] & = 
\begin{pmatrix}
 0 & 1 & 0
 \\
1 & 0 & 0
\\
0 & 0 & 1
\end{pmatrix} \,,
&
V_{\text{pair}}[\sigma_{(13)}] & = 
\begin{pmatrix}
 0 & 0 & 1
 \\
0 & 1 & 0
\\
1 & 0 & 0
\end{pmatrix} \,,
\\
V_{\text{pair}}[\sigma_{(123)}] & = 
\begin{pmatrix}
 0 & 0 & 1
 \\
1 & 0 & 0
\\
0 & 1 & 0
\end{pmatrix} \,,
&
V_{\text{pair}}[\sigma_{(132)}] & = 
\begin{pmatrix}
 0 & 1 & 0
 \\
0 & 0 & 1
\\
1 & 0 & 0
\end{pmatrix} \,,
\end{align}
where $\mathbb{I}_{{S}_{n}}\in {S}_{n}$ is the identity element in ${S}_{n}$ and $\sigma_{(i_{1}\cdots i_{k})}$ is a cycle~\footnote{If $\sigma\in{S}_{n}$ satisfies $\sigma(i_{1})=i_{2}, \sigma(i_{2})=i_{3},\cdots, \sigma(i_{k})=i_{1}$ and $\sigma(i) = i$ for other $i$'s, then this is called a cycle and denoted by $\sigma_{(i_{1},\cdots, i_{k})}$. For example, $\sigma_{(ij)}$ is a permutation between $i$-th and $j$-th elements.}.

Suppose that we use the encoding $U_{\text{E}}(q)=R_{Z}(q_{12})\otimes R_{Z}(q_{13}) \otimes R_{Z}(q_{23})$,
then the induced representation is given by
\begin{align}
V_{\text{pair}}^{\text{(ind)}}[\mathbb{I}_{\mathcal{S}_{3}}] & = \mathbb{I}_{\mathcal{H}_{3}}\,,
&
V_{\text{pair}}^{\text{(ind)}}[\sigma_{(12)}] & = \text{SWAP}_{\llbracket13\rrbracket,\llbracket23\rrbracket}\,,
\\
V_{\text{pair}}^{\text{(ind)}}[\sigma_{(23)}] & = \text{SWAP}_{\dbracket{12},\dbracket{13}}\,,
&
V_{\text{pair}}^{\text{(ind)}}[\sigma_{(13)}] & = \text{SWAP}_{\dbracket{12},\dbracket{23}}\,,
\\
V_{\text{pair}}^{\text{(ind)}}[\sigma_{(123)}] & = 
\text{SWAP}_{\dbracket{12},\dbracket{23}}
\cdot
\text{SWAP}_{\dbracket{12},\dbracket{13}}
\,,
\\
V_{\text{pair}}^{\text{(ind)}}[\sigma_{(132)}] & = \text{SWAP}_{\dbracket{12},\dbracket{23}}\cdot
\text{SWAP}_{\dbracket{13},\dbracket{23}}
\,. &&
\end{align}
where $\text{SWAP}_{\llbracket ij\rrbracket,\llbracket kl \rrbracket}$ is the SWAP operator between the qubits encoding~$q_{ij}$ and~$q_{kl}$ and~$\mathbb{I}_{\mathcal{H}_{n}}$ is the identity operator on the $n$-qubits Hilbert space. 

First, let us consider the $Y$-generators $\mathcal{G}_{Y}=\{Y_{\dbracket{12}}, Y_{\dbracket{13}}, Y_{\dbracket{23}}\}$, where we denote a Pauli matrix $P\in\{I, X, Y, Z\}$ acting on the qubit encoding $q_{ij}$ as $P_{\dbracket{ij}}$.
The twirling formula for this representation leads to
\begin{align}
\mathcal{T}[Y_{\dbracket{12}}]
&=
\frac{1}{6}
\sum_{s\in {S}_{3}} V_{\text{pair}}^{\text{(ind)}}[s] \cdot 
Y_{\dbracket{12}}
V_{\text{pair}}^{\text{(ind)}}[s]^{\dag}
\\
&=
\frac{Y_{\dbracket{12}}+Y_{\dbracket{12}}+Y_{\dbracket{13}}+Y_{\dbracket{23}}
+
Y_{\dbracket{13}}
+
Y_{\dbracket{23}}
}{6} 
\\&=
\frac{1}{3}Y_{\dbracket{12}} + \frac{1}{3}Y_{\dbracket{13}} + \frac{1}{3}Y_{\dbracket{23}} \,,
\\
\mathcal{T}[Y_{\dbracket{13}}]
&=
\frac{1}{6}
\sum_{s\in {S}_{3}} V_{\text{pair}}^{\text{(ind)}}[s] \cdot 
Y_{\dbracket{13}}
V_{\text{pair}}^{\text{(ind)}}[s]^{\dag}
\\
&=
\frac{Y_{\dbracket{13}}+Y_{\dbracket{23}}+Y_{\dbracket{12}}+Y_{\dbracket{13}}
+
Y_{\dbracket{23}}
+Y_{\dbracket{12}}
}{6} 
\\&=
\frac{1}{3}Y_{\dbracket{12}} + \frac{1}{3}Y_{\dbracket{13}} + \frac{1}{3}Y_{\dbracket{23}} \,,
\\
\mathcal{T}[Y_{\dbracket{23}}]
&=
\frac{1}{6}
\sum_{s\in {S}_{3}} V_{\text{pair}}^{\text{(ind)}}[s] \cdot 
Y_{\dbracket{23}}
V_{\text{pair}}^{\text{(ind)}}[s]^{\dag}
\\
&=
\frac{Y_{\dbracket{23}}+Y_{\dbracket{13}}+Y_{\dbracket{23}}
+Y_{\dbracket{12}}
+
Y_{\dbracket{12}}
+Y_{\dbracket{13}}
}{6} 
\\&=
\frac{1}{3}Y_{\dbracket{12}} + \frac{1}{3}Y_{\dbracket{13}} + \frac{1}{3}Y_{\dbracket{23}} \,.
\end{align}
Thus, the symmetrized ansatz can be constructed by
\begin{align}
\exp\left[
i \left(
\theta'_{1} \mathcal{T}[Y_{\dbracket{12}}]
+
\theta'_{2}
\mathcal{T}[Y_{\dbracket{13}}]
+
\theta'_{3}
\mathcal{T}[Y_{\dbracket{23}}]
\right)
\right]
=
\exp\left[
i \theta\left(
Y_{\dbracket{12}}
+Y_{\dbracket{13}}
+Y_{\dbracket{23}}
\right)
\right]\,,
\end{align}
where
\begin{align}
\theta &= \frac{1}{3}\theta'_{1}+\frac{1}{3}\theta'_{2} + \frac{1}{3}\theta'_{3}\,.
\end{align}

Next, let us consider $ZZ$-generators $\mathcal{G}_{ZZ} = \{ Z_{\dbracket{12}}Z_{\dbracket{13}}, Z_{\dbracket{13}}Z_{\dbracket{23}}, Z_{\dbracket{23}}Z_{\dbracket{12}} \}$.
The twirling formula gives
\begin{align}
\mathcal{T}[Z_{\dbracket{12}}Z_{\dbracket{13}}]
& =
\frac{1}{6}
\sum_{s\in {S}_{3}} V_{\text{pair}}^{\text{(ind)}}[s] \cdot 
(Z_{\dbracket{12}}Z_{\dbracket{13}})\cdot
V_{\text{pair}}^{\text{(ind)}}[s]^{\dag}
\\
&=
\frac{1}{6}
\big[
Z_{\dbracket{12}}Z_{\dbracket{13}}
+ Z_{\dbracket{12}}Z_{\dbracket{23}} 
+Z_{\dbracket{13}}Z_{\dbracket{12}}
\notag\\&\quad\quad\quad
+Z_{\dbracket{23}}Z_{\dbracket{13}}
+Z_{\dbracket{13}}Z_{\dbracket{23}}
+Z_{\dbracket{23}}Z_{\dbracket{12}}
\big]
\\&=
\frac{1}{3}
Z_{\dbracket{12}}Z_{\dbracket{13}}
+
\frac{1}{3}
Z_{\dbracket{13}}Z_{\dbracket{23}}
+ \frac{1}{3} Z_{\dbracket{23}}Z_{\dbracket{12}}
\,,
\\
\mathcal{T}[Z_{\dbracket{13}}Z_{\dbracket{23}}]
& =
\frac{1}{6}
\sum_{s\in {S}_{3}} V_{\text{pair}}^{\text{(ind)}}[s] \cdot 
(Z_{\dbracket{13}}Z_{\dbracket{23}})\cdot
V_{\text{pair}}^{\text{(ind)}}[s]^{\dag}
\\
&=
\frac{1}{6}
\big[
Z_{\dbracket{13}}Z_{\dbracket{23}}
+Z_{\dbracket{13}}Z_{\dbracket{23}}
+Z_{\dbracket{12}}Z_{\dbracket{23}}
\notag\\&\quad\quad\quad
+Z_{\dbracket{13}}Z_{\dbracket{12}}
+Z_{\dbracket{23}}Z_{\dbracket{12}}
+Z_{\dbracket{12}}Z_{\dbracket{13}}
\big]
\\&=
\frac{1}{3}
Z_{\dbracket{12}}Z_{\dbracket{13}}
+\frac{1}{3}Z_{\dbracket{13}}Z_{\dbracket{23}}
+ \frac{1}{3} Z_{\dbracket{23}}Z_{\dbracket{12}}
\,,
\\
\mathcal{T}[Z_{\dbracket{23}}Z_{\dbracket{12}}]
& =
\frac{1}{6}
\sum_{s\in {S}_{3}} V_{\text{pair}}^{\text{(ind)}}[s] \cdot 
(Z_{\dbracket{23}}Z_{\dbracket{12}})\cdot
V_{\text{pair}}^{\text{(ind)}}[s]^{\dag}
\\
&=
\frac{1}{6}
\big[
Z_{\dbracket{23}}Z_{\dbracket{12}}
+Z_{\dbracket{13}}Z_{\dbracket{12}}
+Z_{\dbracket{23}}Z_{\dbracket{13}}
+Z_{\dbracket{23}}Z_{\dbracket{12}}
\notag\\&\quad\quad\quad
+Z_{\dbracket{12}}Z_{\dbracket{13}}
+Z_{\dbracket{13}}Z_{\dbracket{23}}
\big]
\\&=
\frac{1}{3}
Z_{\dbracket{12}}Z_{\dbracket{13}}
+
\frac{1}{3}
Z_{\dbracket{13}}Z_{\dbracket{23}}
+ \frac{1}{3} Z_{\dbracket{23}}Z_{\dbracket{12}}
\,,
\end{align}
Thus the equivariant ansatz is obtained as
\begin{align}
&\quad\exp\left[
i \left(
\theta'_{1}\mathcal{T}[Z_{\dbracket{13}}Z_{\dbracket{23}}]
+
\theta'_{2}\mathcal{T}[Z_{\dbracket{13}}Z_{\dbracket{23}}]
+
\theta'_{3}\mathcal{T}[Z_{\dbracket{23}}Z_{\dbracket{12}}]
\right)
\right]\notag
\\&=\exp\left[
i \theta
\left(
Z_{\dbracket{12}}Z_{\dbracket{13}}
+
Z_{\dbracket{13}}Z_{\dbracket{23}}
+ 
Z_{\dbracket{23}}Z_{\dbracket{12}}\right)
\right]\,,
\end{align}
where $\theta=(\theta'_{1}+\theta'_{2}+\theta'_{3})/3$.
\subsection{Self vector}
The representation acting on this space is given by
\begin{align}
V_{\text{self}}[\mathbb{I}_{\mathcal{S}_{3}}] & = 
\begin{pmatrix}
 1 & 0 & 0
 \\
 0 & 1 & 0
 \\
 0 & 0 & 1
\end{pmatrix} \,,
&
V_{\text{self}}[\sigma_{(12)}] & = 
\begin{pmatrix}
0  & 1 & 0
 \\
1 & 0 & 0
\\
0 & 0 & 1
\end{pmatrix} \,,
\\
V_{\text{self}}[\sigma_{(23)}] & = 
\begin{pmatrix}
 1 & 0 & 0
 \\
 0 & 0 & 1
\\
0 & 1 & 0
\end{pmatrix} \,,
&
V_{\text{self}}[\sigma_{(13)}] & = 
\begin{pmatrix}
 0 & 0 & 1
 \\
0 & 1 & 0
\\
1 & 0 & 0
\end{pmatrix} \,,
\\
V_{\text{self}}[\sigma_{(123)}] & = 
\begin{pmatrix}
 0 & 1 & 0
 \\
0 & 0 & 1
\\
1 & 0 & 0
\end{pmatrix} \,,
&
V_{\text{self}}[\sigma_{(132)}] & = 
\begin{pmatrix}
 0 & 0 & 1
 \\
1 & 0 & 0
\\
0 & 1 & 0
\end{pmatrix} \,,
\end{align}
The corresponding induced representation is obtained as
\begin{align}
V_{\text{pair}}^{\text{(ind)}}[\mathbb{I}_{{S}_{3}}] & = \mathbb{I}_{\mathcal{H}_{3}}\,,
&
V_{\text{pair}}^{\text{(ind)}}[\sigma_{(12)}] & = \text{SWAP}_{\llbracket11\rrbracket,\llbracket22\rrbracket}\,,
\\
V_{\text{pair}}^{\text{(ind)}}[\sigma_{(23)}] & = \text{SWAP}_{\dbracket{22},\dbracket{33}}\,,
&
V_{\text{pair}}^{\text{(ind)}}[\sigma_{(13)}] & = \text{SWAP}_{\dbracket{11},\dbracket{33}}\,,
\\
V_{\text{pair}}^{\text{(ind)}}[\sigma_{(123)}] & = 
\text{SWAP}_{\dbracket{22},\dbracket{33}}
\cdot
\text{SWAP}_{\dbracket{11},\dbracket{22}}
\,,
\\
V_{\text{pair}}^{\text{(ind)}}[\sigma_{(132)}] & = \text{SWAP}_{\dbracket{11},\dbracket{33}}\cdot
\text{SWAP}_{\dbracket{11},\dbracket{22}}
\,. &&
\end{align}
Then the twirling formula gives
\begin{align}
\mathcal{T}[Y_{\dbracket{11}}]
&=
\frac{1}{6}
\sum_{s\in {S}_{3}} V_{\text{self}}^{\text{(ind)}}[s] \cdot 
Y_{\dbracket{12}}
V_{\text{self}}^{\text{(ind)}}[s]^{\dag}
\\&=
\frac{1}{3}Y_{\dbracket{11}} + \frac{1}{3}Y_{\dbracket{12}} + \frac{1}{3}Y_{\dbracket{33}} \,,
\\
\mathcal{T}[Y_{\dbracket{22}}]
&=
\frac{1}{6}
\sum_{s\in {S}_{3}} V_{\text{self}}^{\text{(ind)}}[s] \cdot 
Y_{\dbracket{22}}
V_{\text{self}}^{\text{(ind)}}[s]^{\dag}
\\&=
\frac{1}{3}Y_{\dbracket{11}} + \frac{1}{3}Y_{\dbracket{12}} + \frac{1}{3}Y_{\dbracket{33}} \,,
\\
\mathcal{T}[Y_{\dbracket{33}}]
&=
\frac{1}{6}
\sum_{s\in {S}_{3}} V_{\text{self}}^{\text{(ind)}}[s] \cdot 
Y_{\dbracket{33}}
V_{\text{self}}^{\text{(ind)}}[s]^{\dag}
\\&=
\frac{1}{3}Y_{\dbracket{11}} + \frac{1}{3}Y_{\dbracket{22}} + \frac{1}{3}Y_{\dbracket{33}} \,.
\end{align}
Thus the equivariant ansatz is
\begin{align}
\exp\left[
i \left(
\theta'_{1} \mathcal{T}[Y_{\dbracket{11}}]
+
\theta'_{2}
\mathcal{T}[Y_{\dbracket{22}}]
+
\theta'_{3}
\mathcal{T}[Y_{\dbracket{33}}]
\right)
\right]
=
\exp\left[
i \theta\left(
Y_{\dbracket{11}}
+Y_{\dbracket{22}}
+Y_{\dbracket{33}}
\right)
\right]\,.
\end{align}

\section{Performances with alternative settings}
\label{appendix:alternative-settings}
In Section~\ref{sec:results}, we used two layers of hardware efficient ansatz for both the baseline and the rotational invariant QNN. In this appendix, we increase the number of layers to four and also train for $400$ iterations to investigate the performances of more expressive models. Since the fully symmetric model performs well in Section~\ref{sec:results}, we kept it in the same configuration. 

\begin{figure}[H]
     \centering
     \begin{subfigure}[b]{0.45\textwidth}
         \centering
         \includegraphics[width=\textwidth]{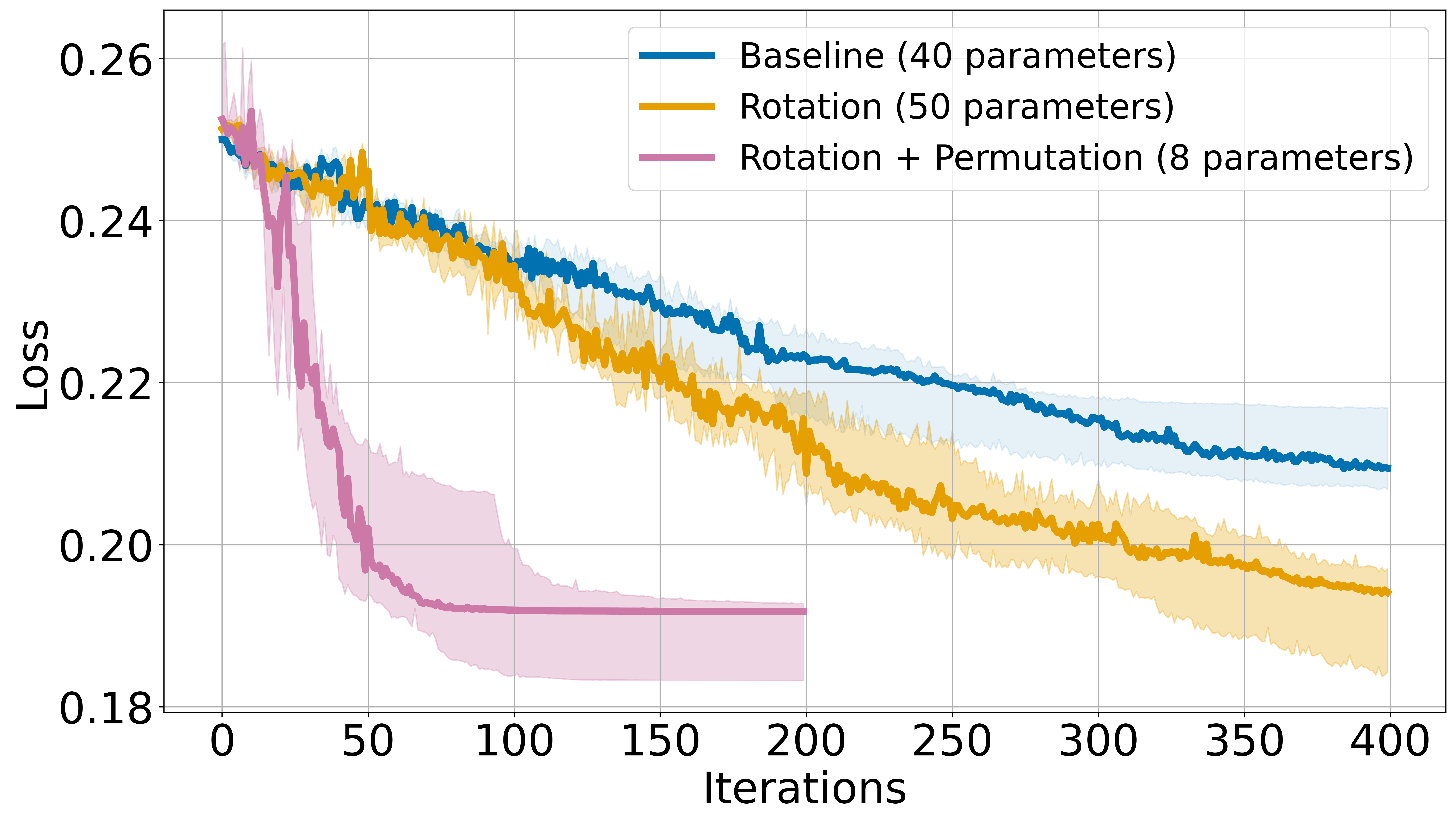}
         \caption{Loss function vs iterations.}
         \label{fig:loss_2dex}
     \end{subfigure}
     \hfill
     \begin{subfigure}[b]{0.45\textwidth}
         \centering
         \includegraphics[width=\textwidth]{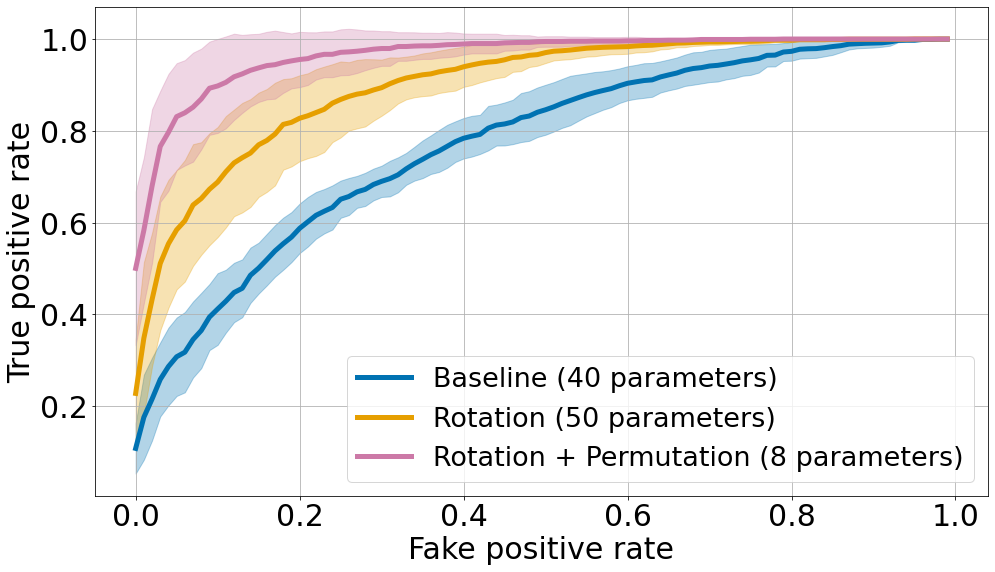}
         \caption{ROC curve.}
         \label{fig:ROC_2dex}
     \end{subfigure}

        \caption{The Loss function during the training process and the ROC curve for various QNN models in the 2D image classification task. The fully symmetric model is the same as the one presented in Section~\ref{ssec:2Drot}. The baseline and the rotationally invariant models have four layers of hardware-efficient ansatz and are trained for $400$ iterations. The loss values shown in the figure are the median values of all ten initializations, whereas the error bands are computed at $25\%$ and $75\%$ quantiles. The ROC curves are obtained by averaging the training results and the error bands are obtained by taking the standard deviations.}
        \label{fig:loss_ROC_2dex}
\end{figure}

We consider the same task of 2D image classification presented in Section~\ref{ssec:2Drot}. The training loss function is shown in Figure~\ref{fig:loss_2dex} and the ROC curve is shown in Figure~\ref{fig:ROC_2dex}. We observe that the performances of the baseline and the rotation-only model are about the same as those listed in Table~\ref{tab:2D}. The extra layers in the ansatz increase the model complexities and thus require mode iterations for the model to converge to the optimized solution. It could be argued that the performance of the two models might be on par with that of the fully symmetric model if it has a sufficient number of layers and was trained for large numbers of iterations far beyond the scope of this experiment. However, given the limited computational resources, we claim it is sufficient to observe a clear advantage in the fully symmetric model since it yields the best performance with a much faster convergence speed due to the lower number of parameters.

\begin{table}[h!]
\centering
\begin{tabular}{ |c|c|c|c| } 
\hline
Methodology & Baseline & Rotation & Rotation + Permutation \\
\hline
$\#$ parameters & 40 & 50  & 8 \\ 
\hline
$\#$ qubits & 8 & 10 & 10\\
\hline
Depth & 4 & 4 & 12\\
\hline
AUC & $0.772 \pm 0.033$ & $0.901 \pm 0.039$ & $0.966 \pm 0.030$\\
\hline
\end{tabular}
\caption{Specifications and performances of the QNNs used in the 2D image classification task. The fully symmetric model is the same as the one presented in Section~\ref{ssec:2Drot}. The baseline and the rotationally invariant models have four layers of hardware-efficient ansatz and are trained for $400$ iterations.}
\label{tab:2Dex}
\end{table}

\twocolumngrid
\bibliography{main}

\end{document}